\documentclass[letterpaper, 10 pt, conference]{ieeeconf}  
\IEEEoverridecommandlockouts                          
          \usepackage{url}              
\usepackage{graphics}
\usepackage{multirow}
\usepackage{amsmath} % assumes amsmath package installed
\usepackage{amssymb}  % assumes amsmath package installed
\usepackage{graphicx}
\usepackage[normalem]{ulem}
\usepackage{subcaption}
\usepackage{hyperref}
\usepackage{algorithm}
\usepackage{algpseudocode}
\usepackage{cite}
\usepackage{soul}
\usepackage{comment}
\usepackage{tikz}
\usepackage{pgfplots}
\usepackage{pgfplotstable}
\usepackage{graphicx}
\usepackage{graphviz}
\usetikzlibrary{math}
\usetikzlibrary{shapes,arrows}
\usetikzlibrary{arrows.meta}
\tikzset{>={Latex[width=2.5mm,length=2.5mm]}}
\usepackage{makecell}

\usetikzlibrary{positioning}
\tikzstyle{block}=[draw opacity=0.7,line width=1.4cm]
\tikzset{arrow_e/.style = {->,> = latex'}}

\newcommand{\vect}[1]{\mathbf{#1}}
\newcommand{\boxend}{\hfill \ensuremath{\Box}}
\newtheorem{thm}{Theorem}[section]

\newtheorem{defn}{Definition}

\newtheorem{prob}{Problem}
\newtheorem{assump}{Assumption}
\allowdisplaybreaks
\title{\LARGE \bf
Microgrids optimal radial reconfiguration via FORWARD algorithm
}
\author{
\\Joan Vendrell, Russell Bent and Solmaz Kia, \emph{Senior Member, IEEE}% <-this % stops a space
\thanks{The first and third authors are with the Mechanical and Aerospace Engineering department,
        University of California Irvine, 
        {\tt\small jvendrel,solmaz@uci.edu}. The second author is with the Theoretical Division,
        Los Alamos National Lab
        {\tt\small rbent@lanl.gov}. }%
}
\allowdisplaybreaks
\newcommand{\longthmtitle}[1]{\mbox{}\textit{{(#1):}}}
\usepackage{placeins}
\begin{document}
\maketitle
%%%%%%%%%%%%%%%%%%%%%%%%%%%%%%%%%%%%%%%%%%%%%%%%%%%%%%%%%%%%%%%%%%%%%%%%%%%%%%%%
\begin{abstract}
Microgrids offer a promising paradigm for integrating distributed energy resources, bolstering energy resilience, and reducing the impact of blackouts. However, their inherent decentralization and dynamic operation present substantial energy management complexities. These complexities, including balancing supply and demand, ensuring system stability, and minimizing operational costs, often necessitate solving computationally intractable NP-hard Mixed-Integer Non-Linear Programming (MINLP) problems. Traditional MINLP solvers struggle with the scalability and feasibility guarantees required for these challenges. To address this, this paper tackles the problem of resource allocation and radial configuration design for microgrid power distribution and proposes and abstracted problem which is solved by introducing a permutation-based iterative search method over the recently introduced \textsf{FORWARD} method to efficiently identify feasible, near-optimal radial network structures while inherently respecting physical constraints. Furthermore, this paper investigates the integration of the proposed method as a warm-start strategy for benchmark MINLP solvers offering a scalable solution for comprehensive microgrid design.

\end{abstract}

\section{Introduction}
\label{sec::intro}
The smart grid paradigm is fundamentally reshaping the energy landscape, driving a critical transition from centralized generation to highly decentralized and distributed architectures. This transformation unlocks significant operational flexibility, enhances grid resilience, and is essential for the deep integration of renewable energy sources and microgrids~\cite{AP-SP-ST-VP-VP:24}. However, this shift to a decentralized topology introduces unprecedented challenges in operational planning and management. Ensuring stability, balancing supply and demand in real-time, and optimizing costs across a complex network of distributed generators are formidable tasks. At the heart of these operational challenges lie complex optimization problems. Their computational intractability stems from the coupling of discrete decisions (e.g., unit commitment, network switching) with the non-linear, non-convex constraints imposed by power flow physics. The performance of existing optimization methods often degrades rapidly with network scale, struggling to guarantee optimality or even find feasible solutions within the stringent time limits required for real-time grid operation. This creates a critical bottleneck for realizing the full potential of the smart grid.

\begin{figure}[t]
    \centering
    \includegraphics[width=0.45\textwidth]{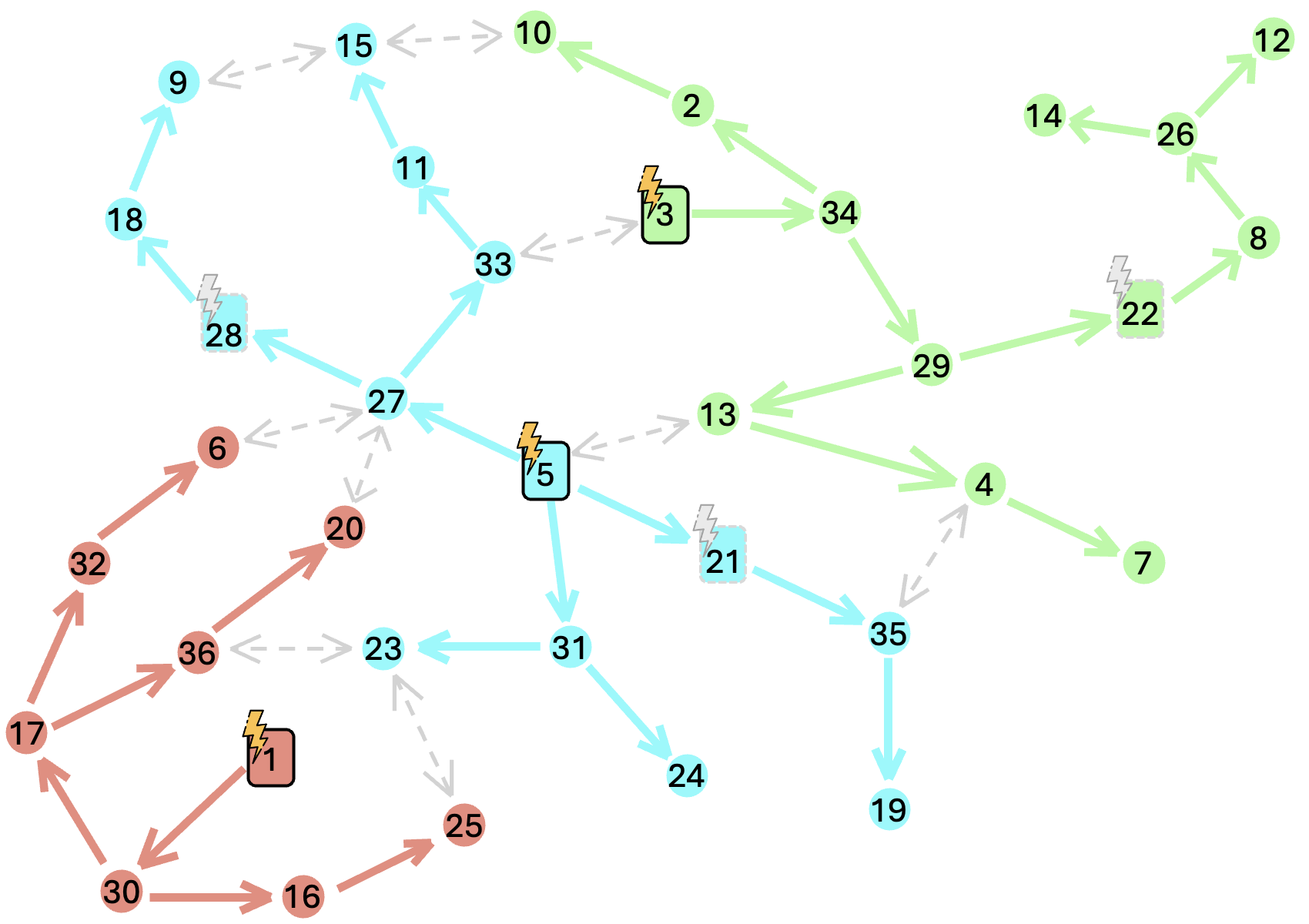}
    \caption{{\small A sample radial configuration for the IEEE 37 network~\cite{SA-SI-AA:22}. In this example, the network is partitioned into three trees rooted at active generators $\{1, 3, 5\}$ (colored red, green, and blue, respectively). Inactive lines and inactive generators $\{21, 22,28\}$ are colored gray.}}
    \label{fig:ieee84}
\end{figure}

In this paper, we address this computational bottleneck by developing a framework for seeking a polynomial-time solution to the optimal power flow and distribution problem within a microgrid. Specifically, given a set of potential generator nodes, we aim to determine both the optimal subset of generators to activate (resource allocation) and the corresponding radial distribution paths to efficiently supply the network's loads; see Fig.~\ref{fig:ieee84} for illustration.

Our objective is to minimize total power losses across the microgrid, a critical focus given that line losses can account for $15-20\%$ of energy distribution in AC systems~\cite{EIA2024}. This optimization is performed while adhering to operational constraints and, crucially, enforcing a radial network topology for simplified control and protection. This inherent combination of discrete (generator/path selection) and continuous (power flow) variables is precisely what renders the problem so computationally intensive.

The problem of interest can be formally cast as a Mixed-Integer Non-Linear Program (MINLP)~\cite{SL-US-MM:11}. These MINLP formulations are inherently NP-hard due to the discrete resource allocation and the complex, non-convex physics of power flow. A classical approach to solving MINLP problems is the Branch-and-Bound method~\cite{YZ-NS-CN-GR:20}. While theoretically capable of finding near-optimal solutions, its search space can grow exponentially with the problem size, leading to intractably large computation times for realistic microgrids. To illustrate the scale of this problem: for a microgrid with $n_g = 10$ potential generators and $\kappa = 3$ active generators, the combinatorial space contains $\binom{10}{3} = 120$ possible configurations. While this seems manageable, for realistic scenarios with $n_g = 50$ and $\kappa = 10$, the space explodes to over $10^{10}$ combinations, making exhaustive search infeasible. Furthermore, this intractability is compounded because, for each one of these configurations, the problem must also find the minimum-loss radial configuration, a search that involves its own set of mixed-integer decisions for edge selection. Some research has explored leveraging graph theory to cleverly modify constraint inequalities~\cite{MK-MP-RR-MS:22} in an attempt to improve scalability. However, these approaches still face significant limitations in handling the complexity and scale of modern distribution networks, motivating the need for the novel polynomial-time approach developed herein.

To overcome the computational intractability and poor scalability of MINLP solvers, a large body of work has explored meta-heuristic and learning-based approaches. These include evolutionary algorithms such as differential evolution~\cite{AT-AAL-MB:17} and genetic algorithms~\cite{CW-YG:13}, as well as artificial immune systems~\cite{GA-RA-ACS-ZZ-WF:22} and particle swarm optimization~\cite{RP-ZN-YM-CC:19}. More recently, machine learning models~\cite{ RB-EB-AM-RR:21} and fuzzy logic methods~\cite{MS-DD-SP:16, JV-SC-VP-GC:20} have been applied to these complex optimization challenges. While these methods can often find high-quality solutions quickly, their primary drawback is the lack of formal guarantees on optimality or, more critically, feasibility. For real-world grid operation, a solution that violates voltage limits or line constraints is operationally unusable, making the reliance on stochastic, non-verifiable methods a significant risk~\cite{MS-DD-SP:16}.

Another research direction attempts to achieve faster solutions by decoupling the problem's discrete and continuous components. For instance, greedy algorithms have been used to select the set of active generators~\cite{JQ-IY-RR:16}, while the problem of finding a radial distribution topology has been approached separately as a minimum spanning tree (MST) problem~\cite{AK-GY-SB-EN-MCC-TL-EP:18} or via branch exchange techniques~\cite{EP-CB-JV:23}. This decomposition strategy, however, introduces a fundamental flaw: the sub-problems are not independent. An "optimal" radial topology found by an MST algorithm, which is blind to power flow physics, may be entirely infeasible once the non-linear power flow constraints are applied. By treating the discrete and continuous optimizations separately, these methods often fail to find a co-dependent, physically valid solution.

\subsection*{Statement of Contribution}

This paper overcomes the trade-off between computational tractability and guaranteed feasibility, which is a key gap in the existing literature, by introducing a novel, two-stage optimization framework based on \emph{hierarchical decomposition}. Our approach decomposes the problem into two levels: a \emph{master problem} that searches over discrete generator combinations, and a \emph{sub-problem} that solves the optimal power flow and topology for each candidate generator set. 

First, we address the sub-problem (optimal power flow and radial topology for a fixed generator set) by reformulating it as a set value optimization problem. This reformulation is the key to unlocking a polynomial-time solution path, as it enables us to leverage the recently developed \textsf{FORWARD} algorithm~\cite{JV-RB-SK:25arxiv}\footnote{Conference version of this paper is available at~\cite{JV-RB-SK:25}. }. We introduce a novel abstraction of the microgrid network that enables the integer variables (network topology) and continuous variables (power flow) to be jointly optimized, thereby avoiding the feasibility failures endemic to decoupled methods. The \textsf{FORWARD} adaptation serves as a polynomial-time \emph{oracle}—a subroutine that efficiently returns high-quality, feasible solutions for any given generator~set.

Second, we address the master problem (optimal resource allocation) using a novel Markov Chain Monte Carlo (MCMC) search strategy. Rather than exhaustively enumerating all $\binom{n_g}{\kappa}$ possible generator combinations, our stochastic local search explores the combinatorial space intelligently, with theoretical guarantees on the probability of finding the optimal solution. We prove that the mixing time of this Markov Chain is polynomial, specifically $\mathcal{O}(\kappa \cdot n_g \cdot \log(n_g))$, ensuring that our framework scales gracefully to large networks.

The key contributions of this unified framework are:
\begin{itemize}
    \item A novel reformulation of the microgrid problem as a set value optimization, enabling the first polynomial-time algorithmic framework to guarantee feasible solutions for this problem class.
    \item A new hierarchical Markov Chain-based method to solve the coupled two-stage problem of resource allocation and network configuration, with rigorous theoretical analysis of convergence via mixing time bounds.
    \item Demonstrated scalability to utility-scale networks (e.g., 8500+ buses), achieving solutions in seconds where traditional MINLP solvers require hours or fail to converge.
    \item The introduction of our framework as a highly effective warm-start strategy, enabling commercial solvers to find optimal solutions for problems that were previously intractable.
\end{itemize}

The remainder of this paper is organized as follows. Section~\ref{sec::preliminary} formally defines the two-stage optimization problem and our network abstraction framework. Section~\ref{sec::power_flow} details the specific power flow physics and engineering constraints. Section~\ref{sec:methodology} presents the proposed hierarchical methodology, detailing both the \textsf{FORWARD}-based inner loop and the Markov Chain-based outer loop. Section~\ref{sec:numeric} describes the experimental setup and presents our numerical results, including scalability analysis and warm-start comparisons. Finally, Section~\ref{sec:conclusions} provides concluding remarks.
%%%%%%%%%%%%%%%%%%%%%%%%%%%%%%%%%%%%%%%%

\section{Problem Definition}
\label{sec::preliminary}
\begin{comment}
\begin{figure*}[htbp]
  \centering
  \begin{subfigure}[b]{0.45\textwidth}
    \centering
    \includegraphics[width=0.6\textwidth]{Fig/ieee18_origin.png}
    \caption{IEEE $18$ distribution network.}
    \label{fig:img1}
  \end{subfigure}
  \hfill
  \begin{subfigure}[b]{0.45\textwidth}
    \centering
    \includegraphics[width=0.7\textwidth]{Fig/ieee18_simple.png}
    \caption{IEEE $18$ abstracted graph.}
    \label{fig:abstraction}
  \end{subfigure}
  \caption{Example of the abstraction process for IEEE $18$ distribution network. Despite its characteristics, energy generators are set as source nodes (square nodes) and energy receivers are set as load nodes (circular nodes). Note that in (a), node $3$ has two different sources available which are unified into a single source in (b). For clarity of presentation, each connection of the three-phase scheme is represented as a single line. }
  \label{fig:sidebyside}
\end{figure*}
\end{comment}
Consider a three-phase alternating current (AC) distribution network, denoted by the graph $\mathcal{G}_D = \mathcal{G}(\mathcal{V}_D, \mathcal{E}_D)$. This network consists of $|\mathcal{V}_D| = N$ nodes (buses) and $|\mathcal{E}_D| = m$ directed edges. Each edge between any two nodes $i,j\in\mathcal{V}_D$ can accommodate multiple physical distribution lines. Specifically, there can be $K$ distinct distribution lines, where each individual line is denoted $(i,j,k)$ with $k\in\{1,2,\cdots,K\}$. Each line $(i,j,k)$ is associated with a set of phases $\Phi_{ijk}\subseteq\{a,b,c\}$, where each phase is represented by $\phi\in\Phi_{ijk}$.

The network $\mathcal{G}_D$ is inherently bidirectional, meaning power can potentially flow in either direction along any edge. However, once a distribution policy (radial configuration) is determined, the actual power flow direction becomes fixed for each active edge. For a given bus $i \in \mathcal{V}_D$, the set of incoming edges is denoted by $\mathcal{E}^+(i)$, the set of outgoing edges is denoted by $\mathcal{E}^-(i)$ and we use $\mathcal{E}^0(i)$ for the set of inactive edges connected to $i$. With some abuse of notation, we use $(i,j,k) \in \mathcal{E}_D$ to denote distribution line $k$ along edge $(i,j) \in \mathcal{E}_D$, where the edge $(i,j)$ represents the potential connection between buses $i$ and $j$, and $k$ identifies the specific physical line within that connection.

Let $\mathcal{V}_g \subset \mathcal{V}_D$ be a set of $n_g$ generator nodes. The remaining $n_c = N - n_g$ nodes constitute the set of \emph{consumers} and \emph{pass-through} nodes, denoted by $\mathcal{V}_c = \mathcal{V}_D \setminus \mathcal{V}_g$. 
Each generator node $i \in \mathcal{V}_g$ has a specified power generation, and each consumer node $j \in \mathcal{V}_c$ has a specified power demand, which can be $0$ when the node acts as a pass-through node. The objective is to determine a \emph{radial configuration} of the distribution network that delivers power from the generator nodes to the consumer nodes to meet the demands while minimizing the overall energy loss across the distribution~lines. 

\begin{defn}[Set of Radial Configurations]
    \label{def::radial}
    A radial configuration of the distribution network $\mathcal{G}_D = \mathcal{G}(\mathcal{V}_D, \mathcal{E}_D)$ with roots at the generator nodes $\mathcal{V}_g \subseteq \mathcal{V}_D$ is a spanning forest subgraph of $\mathcal{G}_D$. This forest must include all nodes in $\mathcal{V}_D$ and have each connected component rooted at, at least, one node in $\mathcal{V}_g$. We denote the set of all possible radial configurations of $\mathcal{G}_D$ with roots at $\mathcal{V}_g$ as $\mathcal{F}(\mathcal{G}_D, \mathcal{V}_g)$. For brevity, when clear from context, we will use only $\mathcal{F}$
    \footnote{A forest graph is an acyclic graph, meaning it contains no cycles. Consequently, a forest is a collection of one or more disjoint trees. \cite{forests}.}.\boxend
\end{defn}

Specifically, in this paper we will consider two problems:
   \begin{prob}\longthmtitle{Optimal Network Radial Reconfiguration under Fixed Generation}\label{prob::Problem1}Given a set of fixed generator nodes $\mathcal{V}_g$ with known power outputs, determine the radial network configuration that minimizes power loss.\boxend\end{prob}
   \begin{prob}\longthmtitle{Optimal Network Radial Reconfiguration and Generator Selection}\label{prob::Problem2} Given a set of generator nodes $\mathcal{V}_g$ with a total power output exceeding demand, select a subset of prespecified $\kappa\in\mathbb{Z}_{\geq 1}$ generators to activate and determine the radial network configuration that minimizes power loss.\boxend
\end{prob}
In what follows, we provide the detailed mathematical model for both Problem~\ref{prob::Problem1} and Problem~\ref{prob::Problem2}. As established in Section~\ref{sec::intro}, casting these problems directly as MINLPs renders them numerically intractable for realistically-sized networks. To overcome this computational barrier, this paper introduces the novel solution approach that forms our core contribution: a polynomial-time framework based on our recently proposed \textsf{FORWARD} algorithm~\cite{JV-RB-SK:25arxiv}. This methodology is rooted in exploring radial graph topological restructuring, informed by the underlying power flow dynamics and operational constraints, to find feasible and high-quality solutions without the exponential complexity of traditional solvers.

%%%%%%%%%%%%%%%%%%%
%%%%%%%%%%%%%%%%%%%
%%%%%%%%%%%%%%%%%%%%%

\section{Mathematical Formulation of the Power Flow Problem}
\label{sec::power_flow}

In this section, we develop the detailed mathematical formulation for the optimization problems introduced in Section~\ref{sec::preliminary}. We first define the necessary notation for a three-phase AC power system. We then formally present the complete MINLP formulation for Problem~\ref{prob::Problem1} (Optimal Reconfiguration). Finally, we extend this formulation to model Problem~\ref{prob::Problem2}  (Joint Generator Selection and Reconfiguration). This complete formulation, which is inherently NP-hard, serves as the computational benchmark that the novel methodology of this paper is designed to overcome.

\subsection{Notation and Physical Properties}
\label{sec::notation}

We introduce the necessary notation for AC distribution networks. Let us define the complex power flowing in an arbitrary distribution line $(i,j,k)$ and phase $\phi\in\Phi_{ijk}$ as 
$$S_{ijk}^\phi = P_{ijk}^\phi + j Q_{ijk}^\phi, \quad k\in\{1,\cdots,K\},$$
where $P_{ijk}^\phi$ and $Q_{ijk}^\phi$ represent the real and imaginary components. Analogously, the admittance on distribution line $(i,j,k)$ is
$$Y_{ijk}^{\phi} = G_{ijk}^\phi +j B_{ijk}^\phi,\quad k\in\{1,\cdots,K\}$$
where $G_{ijk}^\phi$ is the conductance and $B_{ijk}^\phi$ is the susceptance, inversely related to impedance $Z_{ijk}^{\phi}$. By defining $R_{ijk}^{\phi}$ as the resistance and $X_{ijk}^{\phi}$ as the reactance, we have
\begin{equation*}
 \begin{split}
     G_{ijk}^{\phi} = \frac{R_{ijk}^{\phi}}{R_{ijk}^{\phi^2}+X_{ijk}^{\phi^2}}, \quad B_{ijk}^{\phi} = -\frac{X_{ijk}^{\phi}}{R_{ijk}^{\phi^2}+X_{ijk}^{\phi^2}}.
 \end{split}
\end{equation*}

The voltage on a bus $i$ for phase $\phi$ is defined as 
$$V_i^\phi = v_i^\phi\angle\theta_i^\phi$$
where $v_i^\phi$ is the magnitude and $\angle(\cdot)$ denotes the phase angle. The current flowing in distribution line $(i,j,k)$ and phase $\phi\in\Phi_{ijk}$ is defined by Ohm's Law as
\begin{equation*}
 \begin{split}
     \mathbf{x}_{ijk}^\phi &= \sum_{\hat{\phi}\in\Phi_{ijk}} (y_{ijk}^{\phi\hat{\phi}}\angle\theta_{ijk})\cdot(v_i^{\hat{\phi}}\angle\theta_i^{\hat{\phi}}-v_j^{\hat{\phi}}\angle\theta_j^{\hat{\phi}}),
 \end{split}
\end{equation*}
where $y_{ijk}^{\phi\hat{\phi}}$ stands for the magnitude of the admittance and $\theta_{ijk}$ for its angle.

\subsection{MINLP Formulation for Optimal Reconfiguration (Problem 1)}
\label{sec::minlp_formulation_p1}

Using the notation above, we can formally cast the Optimal Network Radial Reconfiguration problem (Problem~\ref{prob::Problem1}) as a Mixed-Integer Non-Linear Program (MINLP). The discrete decisions involve selecting the set of active lines, $\mathcal{S}$, which is encoded using binary variables $\gamma_{ijk} \in \{0, 1\}$ for each potential line $(i,j,k) \in \mathcal{E}_D$. Here, $\gamma_{ijk}=1$ if the line is active and $\gamma_{ijk}=0$ otherwise.

The objective is to minimize the total active power loss, which is formulated as:
\begin{equation}
     \min_{\gamma,\mathbf{x}} \sum_{(i,j,k)\in\mathcal{E}_D}\sum_{\phi\in\Phi_{ijk}} \gamma_{ijk}\cdot R_{ijk}^{\phi}\cdot |\mathbf{x}_{ijk}^\phi|^2\cdot \cos^2{\theta_{ijk}}
     \label{eqn::minlp_objective}
\end{equation}
subject to the following constraints:
\begin{subequations}\label{eq::constraints}
 \small
 \begin{align}
%%%%%%%%%%%%%%%%%%%%%%%%%%%%%%%%%%%%%%%%
% TOPOLOGY
& \gamma \in \mathcal{F}(\mathcal{G}_D, \mathcal{V}_g) \label{const::radiality} \\
%%%%%%%%%%%%%%%%%%%%%%%%%%%%%%%%%%%%%%%%
% PHYSICS
& S_i^{\phi} \!=\!\!\! \sum_{(i,j,k)\in\mathcal{E}^{+}(i)}\!\!\!\! S_{ijk}^{\phi} \!-\! \sum_{(j,i,k)\in\mathcal{E}^{-}(i)}\!\!\!\! S_{jik}^{\phi}, \label{const::balance}\\
& \mathbf{x}_{ijk}^\phi\!=\!\!\sum_{\hat{\phi}\in\Phi_{ijk}}\!\! (y_{ijk}^{\phi\hat{\phi}}\angle\theta_{ijk})\!\cdot\!(v_i^{\hat{\phi}}\angle\theta_i^{\hat{\phi}} \!-\!v_j^{\hat{\phi}}\angle\theta_j^{\hat{\phi}}),  \label{const::ohm} \\
%%%%%%%%%%%%%%%%%%%%%%%%%%%%%%%%%%%%%%%%
% ENGINEERING LIMITS
& \underline{v}_i \leq v_i^\phi \leq \bar{v}_i, \label{const::capacity_v}\\
& P_{ijk}^\phi \leq \gamma_{ijk} \cdot \bar{P}_{ijk}, \label{const::capacity_p}\\
& Q_{ijk}^\phi \leq \gamma_{ijk} \cdot \bar{Q}_{ijk}, \label{const::capacity_q}\\
& \underline{P}_i \leq P_{i}^\phi \leq \bar{P}_{i}\cdot\alpha, \label{const::capacity_p_node}\\
& \underline{Q}_i \leq Q_{i}^\phi \leq \bar{Q}_{i}\cdot\alpha, \label{const::capacity_q_node}\\
& \underline{\theta}\cdot(1\!-\!\gamma_{ijk})\!+\!\underline{\theta}_{ijk} \!\leq\! \angle (V_i^\phi\cdot V_j^{\phi^*}) \!\leq\! \bar{\theta}_{ijk} \!+\! \bar{\theta}\cdot(1\!-\!\gamma_{ijk}) \label{const::phase_angle}
 \end{align}
\end{subequations}
for all $i,j\in \mathcal{V}_D$, $k \in \{1,\dots,K\}$, $(i,j,k)\in\mathcal{E}_D$, and $\phi \in \Phi$. Here, $\gamma$ is the vector of all binary decision variables and $(\cdot)^\star$ stands for the conjugate. Constraint~\eqref{const::radiality} is the topological constraint ensuring the active lines form a radial spanning forest rooted at $\mathcal{V}_g$, as per Definition~\ref{def::radial}. This constraint is itself a combinatorial challenge and is often formulated using additional integer constraints (e.g., a single-commodity flow model~\cite{ML-JF-MR-RR:12}). 

The power flow physics are enforced by~\eqref{const::balance} (Kirchhoff's current law for power balance at each node) and~\eqref{const::ohm} (Ohm's law). The engineering limits are given in~\eqref{const::capacity_v}--\eqref{const::phase_angle}: \eqref{const::capacity_v} bounds nodal voltages; \eqref{const::capacity_p}-\eqref{const::capacity_q} enforce line flow capacities, which are activated only for lines in the configuration (where $\gamma_{ijk}=1$); \eqref{const::capacity_p_node}-\eqref{const::capacity_q_node} limit nodal power injections (where $\alpha$ is a parameter for expected losses, detailed in Section~\ref{sec::preliminary}); and~\eqref{const::phase_angle} bounds the phase angle differences for active lines.

The formulation in~\eqref{eqn::minlp_objective} and \eqref{eq::constraints} is a large-scale, non-convex MINLP. The combination of the combinatorial radiality constraint~\eqref{const::radiality} and the non-linear, non-convex power flow physics (e.g.,~\eqref{eqn::minlp_objective}, \eqref{const::ohm}, \eqref{const::phase_angle}) results in the NP-hard complexity discussed in Section~\ref{sec::intro}. This intractability is the primary motivation for our novel approach, which reformulates this problem to find a polynomial-time solution path.

\subsection{MINLP Formulation for Joint Generator Selection (Problem 2)}
\label{sec::minlp_formulation_p2}

We now extend the formulation to address Problem 2 (Optimal Reconfiguration and Generator Selection Problem). This is a two-stage combinatorial problem where we must select an optimal subset of generators in addition to finding the optimal topology.  To model this, we introduce a new set of binary decision variables, $\beta_i \in \{0,1\}$, for each potential generator node $i \in \mathcal{V}_g$. Here, $\beta_i = 1$ indicates that the generator at node $i$ is activated, and $\beta_i = 0$ indicates it is inactive. The full MINLP for Problem 2 includes all variables $(\gamma, \beta, \mathbf{v}, \mathbf{x})$ and the objective function~\eqref{eqn::minlp_objective}. It also includes all constraints from Problem 1, with the addition of one new constraint and modifications to two existing constraints.

First, we add a cardinality constraint to select exactly $\kappa$ generators, as specified in the problem definition:
\begin{equation}
    \sum\nolimits_{i \in \mathcal{V}_g} \beta_i = \kappa.
    \label{const::cardinality}
\end{equation}
Second, the topological constraint~\eqref{const::radiality} must be modified. The radial forest is no longer rooted at the fixed set $\mathcal{V}_g$, but at the selected set of active generators. This is enforced by:
\begin{equation}
    \gamma \in \mathcal{F}(\mathcal{G}_D, \{i \in \mathcal{V}_g \mid \beta_i = 1\})
    \label{const::radiality_beta}
\end{equation}
This constraint dynamically links the topology decision ($\gamma$) to the generator selection decision ($\beta$). Third, the nodal power injection limits for the potential generator nodes ($i \in \mathcal{V}_g$) must be linked to their activation status. The constraints~\eqref{const::capacity_p_node} and \eqref{const::capacity_q_node} are modified for $i \in \mathcal{V}_g$ as follows:
\begin{subequations}
\small
\begin{align}
    & \underline{P}_i \leq P_{i}^\phi \leq \beta_i \cdot \bar{P}_{i}\cdot\alpha, ~~~~~~~~~~~~~~~~~~~~~\,\forall i\in\mathcal{V}_g,\phi\in\Phi \label{const::capacity_p_node_beta}\\
    & \underline{Q}_i \leq Q_{i}^\phi \leq \beta_i \cdot \bar{Q}_{i}\cdot\alpha, ~~~~~~~~~~~~~~~~~~~~~\forall i\in\mathcal{V}_g,\phi\in\Phi \label{const::capacity_q_node_beta}
\end{align}
\end{subequations}
Here, $\bar{P}_i$ and $\bar{Q}_i$ represent the maximum generation capacity (scaled by $\alpha$) at node $i$, which can only be accessed if $\beta_i = 1$. If $\beta_i = 0$, the injection is limited by its base load, $\underline{P}_i$ and $\underline{Q}_i$. The constraints for all consumer nodes $i \in \mathcal{V}_c$ remain the same as in~\eqref{const::capacity_p_node}-\eqref{const::capacity_q_node}.

This expanded formulation adds a second layer of combinatorial complexity to the already NP-hard MINLP. The solver must now search not only the exponential space of radial topologies but also the combinatorial space of generator subsets. This further highlights the intractability of exact methods for large-scale, two-stage problems and underscores the need for the scalable, polynomial-time framework proposed in this paper.
%%%%%%%%%%%%%%%%%%%%%%%%%%%%%%%%%%%%%%%%%%%%%%%%%
\section{Methodology: A Hierarchical Framework}
\label{sec:methodology}

The previous section detailed the NP-hard MINLP formulations for our two core problems. This section presents our novel, polynomial-time framework designed to overcome this intractability. Our methodology is hierarchical and is presented in two parts, corresponding to the two problems defined in Section~\ref{sec::preliminary}:

\begin{enumerate}
    \item First, we address Problem~\ref{prob::Problem1} (Optimal Reconfiguration), which assumes a fixed set of active generators. We solve this by introducing a sophisticated adaptation of our recently developed \textsf{FORWARD} algorithm~\cite{JV-RB-SK:25arxiv}. This is detailed in Section~\ref{sec:methodology_problem1}.
    
    \item Second, we address the full, two-stage Problem~\ref{prob::Problem2} (Optimal Reconfiguration and Generator Selection). We propose a new framework that leverages our \textsf{FORWARD} adaptation (from Section~\ref{sec:methodology_problem1}) as a high-speed inner-loop solver within a new Markov Chain-based iterative search. This is detailed in Section~\ref{sec:methodology_problem2}.
\end{enumerate}

We begin by detailing the core component of our framework: the adaptation of \textsf{FORWARD} for solving Problem~\ref{prob::Problem1}.

\subsection{Solving Problem 1: Optimal Reconfiguration via FORWARD Adaptation}
\label{sec:methodology_problem1}

The foundational component of our methodology is an adaptation of the \textsf{FORWARD} algorithm~\cite{JV-RB-SK:25arxiv} to solve the fixed-generator Optimal Network Radial Reconfiguration problem (Problem~\ref{prob::Problem1}). The direct application of \textsf{FORWARD} to our multi-phase AC power distribution problem presents several fundamental challenges that require significant algorithmic modifications and careful problem abstraction.

\subsubsection{Challenges in Adapting FORWARD}
\label{sec:methodology_challenges}
The original \textsf{FORWARD} algorithm was designed for abstract flow distribution problems with capacity constraints, formulated as:
\begin{subequations}\label{eqn::problem0}
\begin{align}
(\mathcal{S}^\star,\vect{x}^\star)=&\arg\min ~\sum\nolimits_{(i,j)\in\mathcal{S}}C_{i,j}\cdot x_{i,j}^2,\text{s.t.}\label{eqn::distribution_cost}\\
        &~~~~~~A(\mathcal{S})\vect{x}(\mathcal{S})=\vect{g}-\vect{d},\label{eqn::constraint0}\\
    &~~~~~~0\leq \vect{x}(\mathcal{S})\leq \bar{\vect{x}}(\mathcal{S}),\label{eqn::constraint1}\\
    &~~~~~~ \mathcal{G}(\mathcal{V}_D,\mathcal{S})\in\mathcal{F}(\mathcal{G}_D, \mathcal{V}_g)\label{eqn::radiality}
\end{align}
\end{subequations}
where where $C_{i,j}$ is the cost function coefficient along edge $(i,j)$, $x_{i,j}\in\mathbb{R}_{\geq0}$ is the flow. Constraint~\eqref{eqn::constraint0} represents the linear flow conservation (i.e., Kirchhoff's Current Law), where $\vect{g}$ and $\vect{d}$ are the generation and demand vectors, respectively. Constraint~\eqref{eqn::constraint1} bounds the flow capacity of each edge, and~\eqref{eqn::radiality} enforces a radial topology under multiple fix generation nodes $\mathcal{V}_g$.

Our optimal power distribution problem described by~\eqref{eqn::minlp_objective}-\eqref{eq::constraints} introduces several complexities beyond this abstraction:
\begin{enumerate}
    \item \emph{Multi-phase AC Power Flow}: Unlike single-commodity flow, our problem involves three-phase AC power with complex voltage relationships, phase coupling, and reactive power.
    
    \item \emph{Multi-line Edge Structure}: Each edge $(i,j) \in \mathcal{E}_D$ can contain up to $K$ physical lines $(i,j,k)$ with different electrical characteristics, requiring careful line selection \emph{within} edges.
    
    \item \emph{Complex Constraint Integration}: We must integrate Ohm's law~\eqref{const::ohm}, Kirchhoff's laws~\eqref{const::balance}, voltage limits~\eqref{const::capacity_v}, and phase difference constraints~\eqref{const::phase_angle}.
    
    \item \emph{Non-linear Objective Function}: The power loss objective~\eqref{eqn::minlp_objective} involves quadratic current terms and trigonometric functions.
\end{enumerate}

\subsubsection{Problem Abstraction and Key Assumptions}
\label{sec:methodology_abstraction}
To bridge the gap between the abstract formulation~\eqref{eqn::problem0} and our AC power problem, we first introduce a problem abstraction allowing us to map our complex AC constraints~\eqref{const::ohm}--\eqref{const::phase_angle} into the general framework. Equality constraints \eqref{eqn::constraint0} extends to $h(\cdot)=0$ including Ohm's law and Kirchhoff's laws constraints, while \eqref{eqn::constraint1} is extended to $\underline{g}\leq g(\cdot)\leq \bar{g}$, encapsulating inequality constraints such as voltage limits, phase bounds or capacity constraints. We also define \emph{transmission multi-lines} $\mathcal{L}$ as sequences of consumer nodes connected sequentially from generator nodes in $\mathcal{V}_g$. For any bus $i$, $\mathcal{L}(i)$ denotes the transmission multi-line containing $i$.

To make this reformulation tractable, we leverage two domain-specific properties of power systems. First, total energy loss in distribution networks typically accounts for at most 20\% of the energy supplied. As formalized in \eqref{const::capacity_p_node}--\eqref{const::capacity_q_node}, where generation capacity is adjusted by $\alpha$.

\begin{assump}
Any feasible radial configuration $(\mathcal{V},\mathcal{S}) \subseteq \mathcal{G}_D$ will adhere to energy generation limits, ensuring that the total load served and energy lost will not exceed the total energy available.\boxend
\label{assump::loss}
\end{assump}

Second, voltage magnitudes at all buses are typically maintained within narrow permissible ranges (e.g., 0.95-1.05 p.u.).

\begin{assump}
Using a per-unit representation, we assume $|V_i|=1$ p.u. for all nodes $i\in\mathcal{V}_D$ for the purpose of cost function approximation. Note: The full, non-approximated voltage constraints~\eqref{const::capacity_v} are still enforced for feasibility checks.\boxend
\label{assump::voltage}
\end{assump}

\subsubsection{Key Algorithmic Adaptations for AC Power Flow}
\label{sec:methodology_adaptations}
The \textsf{FORWARD} algorithm (shown in Algorithm~\ref{alg::algorithm}) constructs radial topologies through five key subroutines: \textsf{Pre-Processor}, \textsf{Islander}, \textsf{Net-Concad}, \textsf{Sampler}, and \textsf{Rewire}. To address the challenges identified above, we introduce several critical modifications.

\begin{algorithm}[t]
{\footnotesize
\caption{FORWARD (for Problem 1)}
\begin{algorithmic}[1]
\Require Bidirectional graph $\mathcal{G}_D = \mathcal{G}(\mathcal{V}_D, \mathcal{E}_D)$, Active Generator Set $\bar{\mathcal{V}} \subseteq \mathcal{V}_g$
\State $\mathcal{S} \leftarrow \emptyset$
\State $\mathcal{S},\mathcal{G}_P,\vect{p} \leftarrow$ \textsf{Pre-Processor}($\mathcal{S},\mathcal{G}_D,\vect{p}, \bar{\mathcal{V}}$)
\State $(\mathcal{G}^1,\bar{\mathcal{V}}^1,\vect{p}^1),\cdots,(\mathcal{G}^L,\bar{\mathcal{V}}^L,\vect{p}^L)\leftarrow$\textsf{Islander}($\mathcal{G}_P,\vect{p}, \bar{\mathcal{V}}$)
\For{each partition $\ell\in\{1,\cdots,L\}$}
    \State $\mathcal{S}^\ell\leftarrow \emptyset$
    \State $\mathcal{T}^\ell_i = \mathcal{G}(\{i\},\{\})~~\forall i\in \bar{\mathcal{V}}^\ell$
    \While{$|\mathcal{S}^\ell|\leq |\mathcal{V}^\ell|-1$}
        \State $\bar{\mathcal{G}}^\ell\leftarrow$\textsf{Net-Concad}($\mathcal{G}_D$, $\mathcal{G}^\ell$,$\mathcal{V}^\ell$, $\vect{p}^\ell$)
        \State $e^\star\leftarrow$ \textsf{Sampler}($\bar{\mathcal{G}}^\ell,\vect{p}^\ell,\mathcal{S}^\ell$)
        \State $\mathcal{S}^\ell\leftarrow\mathcal{S}^\ell\cup\{e^\star\}$
        \State Update $\vect{p}^\ell$ and corresponding $\mathcal{T}^\ell_i$
    \EndWhile
    \If{$\mathcal{S}^\ell$ is unfeasible} 
        \State $\mathcal{S}^\ell \leftarrow$ \textsf{Rewire}($\bar{\mathcal{G}}^\ell,\vect{p}^\ell,\mathcal{S}^\ell$)
    \EndIf 
\EndFor
\State $\mathcal{S}\leftarrow \bigcup_{\ell=0}^L\mathcal{S}^\ell$ 
\State \Return Radial configuration $\mathcal{G}(\mathcal{V}(\mathcal{S}),\mathcal{S})$
\end{algorithmic}
\label{alg::algorithm}
}
\end{algorithm}

\begin{enumerate}
\item\emph{Phase-Expanded Node Space}: We expand the node space by representing each physical node $v \in \mathcal{V}_D$ as phase-specific virtual nodes $\{v^\phi\}_{\phi \in \Phi}$, enabling phase-specific demand tracking and proper handling of unbalanced loads.

\item\emph{Multi-line Edge Handling}: For edges with multiple physical lines, we evaluate all lines $(i,j,k)$ within edge $(i,j)$ simultaneously, select the optimal line based on electrical characteristics, and ensure consistent directionality. For each transmission multi-line $\mathcal{L} \in \mathcal{S}$, the current flowing through all buses $\ell \in \mathcal{L}$ remains identical due to the radial structure:
\begin{equation}
\hat{\vect{x}}^{\phi}_{\mathcal{L}} = \sum\nolimits_{\ell\in \mathcal{L}} P_{\ell}^\phi + j Q_{\ell}^\phi
\end{equation}

\item\emph{Physics-Informed Sampling}: Under Assumptions~\ref{assump::loss} and~\ref{assump::voltage}, the cost function~\eqref{eqn::minlp_objective} can be approximated as:
\begin{equation}
 \label{eqn::new_cost}
 \begin{split}
     \hat{f}&(\mathcal{S})= \sum_{(i,j,k)\in\mathcal{S}}\sum_{\phi\in\Phi_{ijk}} R_{ijk}^\phi \cdot |\hat{\vect{x}}^{\phi}_{L}|^2\cdot \cos^2{\theta_{ijk}} \\
     &\!\!=\!\!\! \sum_{(i,j,k)\in\mathcal{S}}\sum_{\phi\in\Phi_{ijk}} \!\!\frac{R_{ijk}^{\phi^3}}{Z_{ijk}^{\phi^2}}\!\cdot\!\Big((\sum_{\ell\in \mathcal{L}(ijk)} P_{\ell}^\phi)^2 \!+\! (\!\sum_{\ell\in \mathcal{L}(ijk)} \!\!Q_{\ell}^\phi)^2\Big).
 \end{split}
\end{equation}
For sequential implementation, we define a surrogate function based on accumulated demand:
\begin{equation}
\label{eqn::surrogate_cost}
 \begin{split}
     \hat{f}(\mathcal{S})\!=\!\!\!\! \!\!\sum_{(i,j,k)\in\mathcal{S}}\sum_{\phi\in\Phi_{ijk}}\!\! \frac{R_{ijk}^{\phi^3}}{Z_{ijk}^{\phi^2}}\!\cdot\!\Big((\!\!\!&\sum_{\ell\in \mathcal{L}^{-}(ijk)}\!\!\!\! P_{\ell}^\phi)^2\!+\! (\!\!\!\!\!\sum_{\ell\in \mathcal{L}^{-}(ijk)}\!\!\! Q_{\ell}^\phi)^2\Big)
 \end{split}
\end{equation}
This enables the submodular optimization framework required for \textsf{FORWARD}. The core innovation is a redesigned \textsf{AC-Sampler} (Alg.~\ref{alg::ac_sampler}) that uses a physics-informed weight function:
\begin{equation}
\label{eqn::distribution_prob2}
w^\phi_{i,j,k} = \frac{p^\phi_i + q^\phi_i}{\sum_{(\ell,\ell-1,\kappa)\in \mathcal{L}^{+}(j)} \frac{R^{\phi}_{\ell,\ell-1,\kappa}}{Z^\phi_{\ell,\ell-1,\kappa}} d^{\phi}_{\ell}}
\end{equation}
where $d^{\phi}_{\ell}$ is the accumulated apparent power demand.

\item \emph{Constraint Integration Strategy}: We embed the complex power system constraints directly within the \textsf{FORWARD} subroutines. The \textsf{Sampler} and \textsf{Rewire} modules actively check constraints~\eqref{const::balance}--\eqref{const::phase_angle} during construction. Edges that would lead to a violation are marked as infeasible and pruned from the search, ensuring that the final radial configuration $\mathcal{S}$ is, by construction, a feasible solution to the full MINLP.
\end{enumerate}

\begin{algorithm}[t]
\footnotesize
\caption{\textsf{AC-Sampler}}
\begin{algorithmic}[1]
\Require Sub-graph partition $\bar{\mathcal{G}}^\ell$, phase-expanded flow vector $\vect{p}^\ell$, current configuration $\mathcal{S}^\ell$
\State Initiate priority queue $q\leftarrow\emptyset$
\State Compute AC power-aware weights $\forall (i,j,k)\in\mathcal{E}^\ell$ using~\eqref{eqn::distribution_prob2}
\State Filter edges violating constraints from~\eqref{const::balance}--\eqref{const::phase_angle}
\State Sort $q$ in increasing order of power loss potential
\State Apply edge-delete procedure and prioritization rules
\State $e^\star \leftarrow \textup{pop from } q$
\State Update phase-expanded node states for all phases $\phi \in \Phi_{ijk}$
\State Update multi-line transmission paths $\mathcal{L}^{+}(i)$ and $\mathcal{L}^{-}(j)$
\State \Return $e^\star$ and associated line selection $(i,j,k)$
\end{algorithmic}
\label{alg::ac_sampler}
\end{algorithm}

\subsection{Solving Problem 2: Resource Allocation via Hierarchical Decomposition}
\label{sec:methodology_problem2}

The \textsf{FORWARD} adaptation detailed in Section~\ref{sec:methodology_problem1} provides a polynomial-time solution for Problem~\ref{prob::Problem1}, but it is predicated on the fact that the active generator set $\bar{\mathcal{V}} \subseteq \mathcal{V}_g$ is fixed. Problem~\ref{prob::Problem2}, however, is a two-stage combinatorial problem that requires us to \emph{select} the optimal subset of $\kappa$ generators from the $n_g$ candidates. We now propose a complete hierarchical framework to solve this.

\subsubsection{A Primal Decomposition Framework}
\label{sec:methodology_decomposition}

To formalize this two-stage problem, we follow a primal decomposition strategy~\cite{geoffrion1970primal}. We reformulate Problem~\ref{prob::Problem2} into a \emph{master problem} and a \emph{decomposed subproblem}. The master problem \eqref{eq::master_problem} optimizes over the discrete, combinatorial set of all possible $\kappa$-combinations of generators:
\begin{align}\label{eq::master_problem}
\min_{\bar{\mathcal{V}} \subset \mathcal{V}_g, |\bar{\mathcal{V}}|=\kappa} f(\bar{\mathcal{V}})
\end{align}
where the cost function $f(\bar{\mathcal{V}})$ is itself the optimal value of a subproblem. This \emph{subproblem} \eqref{eq::decomposed_subproblem} is precisely Problem 1: for a given generator set $\bar{\mathcal{V}}$, find the minimal power loss for the best radial topology rooted at that set:
\begin{subequations}
\begin{align}\label{eq::decomposed_subproblem}
&f(\bar{\mathcal{V}}) = \min_{\mathcal{S} \in \mathcal{F}(\mathcal{G}_D, \bar{\mathcal{V}})} \left( \min_{\vect{x}} \sum_{(i,j,k)\in\mathcal{S}}C_{ijk}(\psi)\cdot x_{ijk}^2 \right) \\
&\qquad\text{subject to constraints}~\eqref{const::balance} - \eqref{const::phase_angle}.
\end{align}
\end{subequations}

This decomposition does not reduce the NP-hard complexity, but it formally clarifies our hierarchical methodology:
\begin{enumerate}
    \item \emph{The Subproblem Solver (Oracle):} The \textsf{FORWARD} adaptation (Algorithm~\ref{alg::algorithm}) acts as our polynomial-time \emph{oracle}. In optimization theory, an oracle is a subroutine that, given an input, returns a solution (not necessarily optimal) in polynomial time. Here, \textsf{FORWARD} provides high-quality feasible solutions for any given generator set $\bar{\mathcal{V}}$ from the master problem.
    
    \item \emph{The Master Problem Solver:} We require an smart search algorithm to explore the master's state space, $\binom{n_g}{\kappa}$, which is too large for brute-force enumeration.
\end{enumerate}

\subsubsection{Stochastic Local Search via Markov Chain}
\label{sec:methodology_sls}

To solve the master problem~\eqref{eq::master_problem}, we propose a Stochastic Local Search (SLS) framework, a class of methods highly effective for large-scale combinatorial optimization~\cite{HHS-book:04}. This approach is implemented as a Markov Chain that ``walks" through the state space of possible generator combinations, as shown in Algorithm~\ref{alg:markov}.

The initialization of $\bar{\mathcal{V}}$ can significantly impact convergence speed. We employ a greedy initialization strategy: we rank generators by their proximity to high-demand nodes and select the top $\kappa$ candidates. However, our theoretical results (Theorem~\ref{thm::markov}) hold for any initialization.

\begin{algorithm}[t]
\footnotesize
\caption{Permutation Search over \textsf{FORWARD} (for Problem 2)}
\label{alg:markov}
\begin{algorithmic}[1]
\Require $\mathcal{G}_D(\mathcal{V}_D, \mathcal{E}_D)$, Source Set $\mathcal{V}_g \subset \mathcal{V}_D$, Cardinality $\kappa$
\Ensure Optimal radial reconfiguration $\mathcal{S}_{\min}$
\State Initialize a set $\bar{\mathcal{V}} \subset \mathcal{V}_g$ with $|\bar{\mathcal{V}}| = \kappa$ \Comment{Initial state for master problem}
\State $\mathcal{S}_{\min}\gets\textsf{FORWARD}(\mathcal{G}_D, \bar{\mathcal{V}})$ \Comment{Call subproblem oracle}
\State Compute $f_{\min} \gets f(\mathcal{S}_{\min})$
\For{$i = 1$ to $t_{\max}$}
  \State $\bar{\mathcal{V}}_{new} \gets \bar{\mathcal{V}}$
  \State Randomly select $v_i \in \bar{\mathcal{V}}_{new}$ and $v_j \in \mathcal{V}_g \setminus \bar{\mathcal{V}}_{new}$
  \State $\bar{\mathcal{V}}_{new} \gets (\bar{\mathcal{V}}_{new} \setminus \{v_i\}) \cup \{v_j\}$ \Comment{Take step in master problem state space}
  \State $\mathcal{S}_i\gets\textsf{FORWARD}(\mathcal{G}_D, \bar{\mathcal{V}}_{new})$ \Comment{Call subproblem oracle}
  \State Compute $f_i \gets f(\mathcal{S}_i)$
  \If{$f_i \leq f_{\min} + \epsilon$} \Comment{Accept new state (greedy)}
    \State $\mathcal{S}_{\min}\gets\mathcal{S}_i$
    \State $f_{\min} \gets f_i$
    \State $\bar{\mathcal{V}} \gets \bar{\mathcal{V}}_{new}$
  \EndIf
\EndFor
\State \Return $\mathcal{S}_{\min}$
\end{algorithmic}
\end{algorithm}

Algorithm~\ref{alg:markov} is a form of Local Search~\cite{UF-VM-JV:11}. It initializes a generator combination $\bar{\mathcal{V}}$ (using the greedy strategy described above) and evaluates it using \textsf{FORWARD}. It then iteratively explores the ``neighborhood" of the current state by ``swapping" one active generator for one inactive generator. This constitutes a greedy random walk, as the new state is only accepted if it improves upon the current best solution. 

The acceptance criterion in line 10 employs a tolerance $\epsilon$ to account for the fact that \textsf{FORWARD} is a heuristic oracle that may not return the exact optimal value of $f(\bar{\mathcal{V}})$. We set $\epsilon = 0.01 \cdot f_{\min}$ (1\% relative tolerance) in our experiments (Section~\ref{sec:numeric}), allowing the search to escape local plateaus while maintaining solution quality.

The iteration limit $t_{\max}$ is set based on Theorem~\ref{thm::markov} (presented below). To achieve a success probability $p \geq 0.95$, we require $t_{\max} \geq 0.95 \cdot \kappa \cdot n_g \cdot (2 + \log(n_g))$. In practice, we use $t_{\max} = \max(\text{max\_iters}, \lceil 0.95 \cdot \kappa \cdot n_g \cdot (2 + \log(n_g)) \rceil)$ where \emph{max\_iters} is a user-specified parameter for early termination.

\subsubsection{Markov Chain Mixing Time Analysis}
\label{sec:methodology_mcmc}

A critical question is how many iterations $t$ are sufficient for this random walk to find the optimal solution. This is answered by the \emph{mixing time} ($t_{\text{mix}}$) of the Markov Chain.

Intuitively, the mixing time is the ``warm-up period" required for the random walk to (asymptotically) forget its starting point and converge to its stationary distribution. A common analogy is shuffling a deck of cards: the mixing time is the number of shuffles required until the deck is considered ``sufficiently random." In our optimization context, we are not seeking a random sample, but rather leveraging this theory. The mixing time provides a theoretical bound on the number of steps needed to have a high probability of \emph{visiting} the optimal state (the best generator set). In our context, the stationary distribution of the Markov Chain is not uniform over all generator sets, but it is biased toward low-loss configurations due to our greedy acceptance rule. This metric tells us how many ``swaps" are needed before we can expect to have visited (and thus identified) the optimal generator set with high probability. Crucially, this mixing time grows only polynomially with problem size, not exponentially.

This MCMC approach is a cornerstone of modern computational science, used for sampling and optimization in fields like statistical physics (e.g., Simulated Annealing~\cite{kirkpatrick1983simulated}), Bayesian statistics, and operations research. Our application, which searches a combinatorial state space, is highly analogous to its use in other NP-hard problems, such as the graph-coloring problem~\cite{gharan2017lecture6}.

The state space of all $\kappa$-combinations can be geometrically represented as a search over a \emph{permutohedron} $\mathcal{P}$~\cite{JB:72}, where each node (a generator set) is connected to all other sets reachable by a single ``swap." This formal structure allows us to bound the mixing time.

\begin{thm}
    \label{thm::markov}
    Given a ground set $\mathcal{V}_D$, a source set $\mathcal{V}_g\subseteq\mathcal{V}_D$ with $n_g$ sources, an initial solution $\bar{\mathcal{V}}$, and $t$ iterations where, at each one, one element of $\bar{\mathcal{V}}$ is changed by an element in $\mathcal{V}_g\setminus\bar{\mathcal{V}}$, the least dissipative combination will be found with probability
    $$p \geq \frac{t}{\kappa\cdot n_g\cdot(2+\log(n_g))}.$$
\end{thm}

\begin{proof}
    We bound the mixing time using the coupling method, a standard technique in Markov Chain analysis. The key insight is to construct two coupled chains $(X_t, Y_t)$ starting from different initial generator sets and show that their expected distance decreases exponentially. The rate of this decrease determines the mixing time.
    
    The proof proceeds by using coupling methods~\cite{RB-MD:97}, a standard technique for bounding the mixing time of a Markov Chain. Let $(X_t,Y_t)$ be a coupling such that for some $\alpha\geq 0$ and $x,y\in\mathcal{V}_g$ with $x\sim y$. By \cite{RB-MD:97}, if it holds
    $$\mathbb{E}[d(X_{t+1},Y_{t+1})|X_t=x,Y_t=y]\leq e^{-\alpha} d(x,y)$$
    then the mixing time is $t_{\text{mix}}(\epsilon)\leq\frac{1}{\alpha}(\log(\text{diam}(\mathcal{P}))+\log(\frac{1}{\epsilon}))$.
    
    The similarity to the graph coloring problem arises because both problems involve selecting a subset of elements %(colors for nodes, generators for activation) 
    from a larger set, where the state space can be represented as a permutohedron. In graph coloring, vertices represent valid colorings, and edges connect colorings differing by one color swap. In our problem, vertices represent valid generator sets, and edges connect sets differing by one generator swap. The coupling analysis for both problems follows the same mathematical structure, allowing us to adapt the mixing time bound from~\cite{gharan2017lecture6}.
    
    Observing the similarity with respect to the classical graph-coloring problem~\cite{gharan2017lecture6}, it is straightforward to see that
    $$\mathbb{E}[d(X_{t+1},Y_{t+1})|X_t=x,Y_t=y]\leq e^{-\frac{t+1}{n_g\cdot\kappa}} n_g$$
    and, consequently, $t_{\text{mix}}\leq \kappa\cdot n_g\cdot(2+\log(n_g))$.
    Finally, if $t$ iterations are done over the Markov Chain, the optimal combination will be observed with probability $p \geq \frac{t}{t_{\text{mix}}}$.
\end{proof}
\medskip

Theorem~\ref{thm::markov} provides the core theoretical justification for our hierarchical framework. It can be interpreted as follows:
\begin{enumerate}
\item  \emph{Probabilistic Guarantee:} It provides a probabilistic guarantee of finding the optimal generator set. The probability of success, $p$, increases linearly with the number of iterations, $t$.
\item  \emph{Cost of Search:} The denominator, $t_{\text{mix}} \approx \mathcal{O}(\kappa \cdot n_g \cdot \log(n_g))$, represents the cost of the search.
\item \emph{Scalability:} Most importantly, this cost is not exponential, but low-order polynomial in the number of potential generators ($n_g$) and the selected ones ($\kappa$).
\end{enumerate}

\noindent\textbf{Computational Complexity of the Complete Framework:} Each iteration of Algorithm~\ref{alg:markov} requires one call to \textsf{FORWARD}, which has complexity $\mathcal{O}(m \cdot N)$ for a network with $m$ edges and $N$ nodes (as established in Section~\ref{sec:methodology_problem1}). With $t_{\max} = \mathcal{O}(\kappa \cdot n_g \cdot \log(n_g))$ iterations, the total complexity is $\mathcal{O}(\kappa \cdot n_g \cdot \log(n_g) \cdot m \cdot N)$. For fixed $\kappa$ and $n_g$ (typical in practice), this reduces to $\mathcal{O}(m \cdot N)$ per iteration, making the framework highly scalable.

In essence, the theorem proves that our stochastic search is scalable, as there is no need of an exponential number of iterations to have a high probability of finding the optimal generator set. This, combined with our polynomial-time \textsf{FORWARD} oracle for the subproblem, results in a complete, hierarchical framework that finds high-quality solutions for the full two-stage problem in polynomial time, overcoming the NP-hard barrier of MINLP formulation. As we demonstrate in Section~\ref{sec:numeric}, this theoretical scalability translates to practical performance: our framework solves Problem~\ref{prob::Problem2} for 8500 buses network in under 3 minutes, while commercial MINLP solvers time out after 78 hours (Table~\ref{tab:cost_b}).

\subsubsection{Practical Implementation Considerations}
\label{sec:methodology_practical}

While Theorem~\ref{thm::markov} provides theoretical guarantees, several practical factors influence performance:

\begin{enumerate}
    \item \emph{Warm Starting:} If a good initial generator set is available (e.g., from historical data), initializing $\bar{\mathcal{V}}$ with this set can significantly reduce convergence time while maintaining the theoretical guarantees.
    
    \item \emph{Parallel Evaluation:} The \textsf{FORWARD} oracle calls in lines 2 and 8 of Algorithm~\ref{alg:markov} for different candidate sets are independent and can be parallelized, enabling significant speedups on multi-core architectures.
    
    \item \emph{Adaptive Tolerance:} The tolerance $\epsilon$ can be gradually tightened as the search progresses (e.g., starting at $\epsilon = 0.05 \cdot f_{\min}$ and reducing to $\epsilon = 0.001 \cdot f_{\min}$), allowing faster initial exploration followed by fine-tuning.
    
    \item \emph{Early Stopping:} If $f_{\min}$ does not improve for a specified number of consecutive iterations %(e.g., $0.1 \cdot t_{\max}$)
    , the search can be terminated early, trading theoretical guarantees for computational speed in time-critical applications.
\end{enumerate}
%%%%%%%%%%%%%%%%%%%%%%%%%%%%%%%%%%%%%%%%%%%%%%%%%
%%%%%%%%%%%%%%%%%%%%%%%%%%%%%%%%%%%%%%%%%%%%%%%%%

%%%%%%%%%%%%%%%%%%%%%%%%%%%%%%%%%%%%%%%%%%%%%%%%%
%%%%%%%%%%%%%%%%%%%%%%%%%%%%%%%%%%%%%%%%%%%%%%%%%
\section{Numerical example}
\label{sec:numeric}

The performance evaluation of the proposed \textsf{FORWARD}-based methodology encompasses an analysis across benchmark networks, real-world systems, and synthetic topologies to demonstrate both computational efficiency and solution quality. This section presents the experimental framework, comparative analysis with established MINLP solvers, and evaluation of the warm-start integration strategy.

Our experimental design addresses the scalability limitations of existing approaches by testing on networks ranging from small IEEE benchmarks to large-scale synthetic systems with up to 8500 buses. The test suite includes standard IEEE benchmark networks (11, 18, 33, 37, and 70 buses) obtained from OpenDSS~\cite{DM-RD-JT-MM:22}, which serve as validation baselines following established academic literature. To bridge the gap between academic benchmarks and real-world applications, we incorporate larger networks including the 84-bus feeder from Taiwan Power Company~\cite{SA-SI-AA:22}, the 136-bus system from Três Lagoas, Brazil~\cite{MM-RR:21}, and the IEEE 8500-bus system representing utility-scale distribution networks.

Additionally, we generate synthetic networks using the Watts-Strogatz model~\cite{DW-SS:98} to emulate the small-world properties characteristic of distribution networks, where few nodes exhibit large connectivity while most maintain small connectivity patterns. Three synthetic networks with 120, 240, and 400 buses are designed to evaluate scalability beyond existing benchmarks\footnote{All data and code are available at~\cite{forwardrepo}.}. 

For microgrid design using the abstracted problem formulation~\eqref{eqn::problem0}, all datasets undergo preprocessing and transformation into graph networks $\mathcal{G}_D$. All energy sources are modeled homogeneously, differing only in generation capacity, while multi-objective optimization considering energy source heterogeneity remains beyond this paper's scope. Similarly, all electrical elements requiring energy (e.g., capacitors) are represented as load nodes. Table~\ref{tab:overview} summarizes the characteristics of all test networks, including node counts, edge densities, active source distributions, and power demand profiles.

\begin{table*}[t]
\centering
\caption{Comprehensive Test Network Characteristics}
\label{tab:overview}
\begin{tabular}{|c|c|c|c|c|c|c|}
\hline
\textbf{Network} & \textbf{Nodes} & \textbf{Edges} & \textbf{Active Sources} & \textbf{Total Demand (kW)} & \textbf{Total Generation (kW)} & \textbf{Network Type} \\ 
\hline
\multicolumn{7}{|c|}{\textit{IEEE Benchmark Networks}} \\
\hline
IEEE 11 & 11 & 24 & 2 & 5,400 & 5,400 & Radial \\
IEEE 18 & 18 & 36 & 2 & 2,993 & 5,000 & Meshed \\
IEEE 33 & 33 & 74 & 3 & 4,973 & 5,520 & Radial \\
IEEE 37 & 37 & 86 & 3 & 3,548 & 16,500 & Unbalanced \\
IEEE 70 & 70 & 146 & 2 & 3,396 & 7,500 & Large Radial \\
\hline
\multicolumn{7}{|c|}{\textit{Real-World Distribution Networks}} \\
\hline
IEEE 84 & 84 & 194 & 2 & 4,335.3 & 27,500 & Taiwan Feeder \\
IEEE 136 & 136 & 308 & 3 & 6,103.9 & 20,000 & Brazilian System \\
IEEE 8500 & 8500 & 5058 & 9 & 11,081.55 & 66,250 & Utility-Scale \\
\hline
\multicolumn{7}{|c|}{\textit{Synthetic Networks (Watts-Strogatz Model)}} \\
\hline
WS 120 & 120 & 560 & 5 & 65,726 & 292,115.6 & Small-World \\
WS 240 & 240 & 1020 & 5 & 131,597 & 438,656.7 & Small-World \\
WS 400 & 400 & 1680 & 9 & 213,204 & 947,573.4 & Small-World \\
\hline
\end{tabular}
\end{table*}

All experiments are implemented using the PowerModelsDistribution~\cite{FOBES2020106664} framework from Los Alamos National Laboratory with Julia on a 32-core Intel(R) Core(TM) i9-14900F with 32 GB RAM. We compare our methodology against two established solvers representing different optimization paradigms: \textsf{JUNIPER} (open-source Branch-and-Bound with deep search trees and NLP relaxations~\cite{OK-CC-HH-HN:18}) and \textsf{KNITRO} (commercial Branch-and-Bound with hybrid optimization methods~\cite{RB-JN-RW:06}). A comprehensive time limit of 78 hours is imposed for all experiments, with "TL" indicating timeout scenarios. Performance evaluation encompasses four critical metrics: energy loss minimization, computational time efficiency, memory allocation requirements, and percentage loss relative to total system capacity.

\subsection{Radial distribution under fixed resource allocation}

%%%%%%%%%%%%%%%%%% COST %%%%%%%%%%%%%%%%%%%
\begin{table}[t]
\centering
\caption{Distribution loss comparison between Knitro, Juniper and \textsf{FORWARD} (in kW). }
\label{tab:cost_a}
\begin{tabular}{|c|c|c|c|}
\hline
\textbf{Network} & \textbf{\textsf{FORWARD}} & \textbf{\textsf{KNITRO}} & \textbf{\textsf{JUNIPER}}  \\ 
\hline
IEEE 11 & 360.183 & 663.561 & 360.183 \\
\hline
IEEE 18 & 175.821 & 204.328  & 499.356 \\
\hline
IEEE 33 & 919.783  & 318.568  & 996.900  \\
\hline
IEEE 37 & 598.720 & 623.54 & TL \\
\hline
IEEE 70 & 461.870  & TL & TL \\
\hline
\hline
IEEE 84 & 423.360 & TL  & TL \\
\hline
IEEE 136 & 981.150 & TL  & TL  \\
\hline
IEEE 8500 & 1,391.90 & TL  & TL  \\
\hline
\hline
WS 120 & 1,428.72 & TL & TL \\
\hline
WS 240 & 4,393.17 & TL & TL \\
\hline
WS 400 & 28,345.7 & TL & TL \\
\hline
\end{tabular}
\end{table}

%%%%%%%%%%%%%%%%%%%% TIME %%%%%%%%%%%%%%%%%%%
\begin{table}[t]
\centering
\caption{CPU time comparison between Knitro, Juniper and \textsf{FORWARD} (in seconds). }
\label{tab:cpu_a}
\begin{tabular}{|c|c|c|c|}
\hline
\textbf{Network}  & \textbf{\textsf{FORWARD}} & \textbf{\textsf{KNITRO}} & \textbf{\textsf{JUNIPER}} \\ 
\hline
IEEE 11 & 0.03274 & 4.18872 & 74.8925\\
\hline
IEEE 18 & 0.22865 & 123.416 & 400.508\\
\hline
IEEE 33 & 0.06617 & 3,345.67 & 17,527.1 \\
\hline
IEEE 37 & 4.05542 & 4,252.43 & TL \\
\hline
IEEE 70 & 0.12199 & TL & TL \\
\hline
\hline
IEEE 84 & 0.17410 & TL & TL \\
\hline
IEEE 136 & 0.36849 & TL & TL \\
\hline
IEEE 8500 & 171.322 & TL & TL\\
\hline
\hline
WS 120 & 0.36053 & TL & TL \\
\hline
WS 240 & 1.01575 & TL & TL \\
\hline
WS 400 & 3.08962 & TL & TL \\
\hline
\end{tabular}
\end{table}

In this section, we show the solutions provided by Algorithm~\ref{alg:markov}, \textsf{KNITRO} and \textsf{JUNIPER} with respect to Problem $1$ where the amount of active sources is fixed and known. Note that under this conditions, Algorithm~\ref{alg:markov} simplifies to the direct implementation of \textsf{FORWARD}. In Table~\ref{tab:cost_a}, we can observe that classical solvers poorly scale with graph size being unable to produce any result for real-world networks. 

It is also remarkable the capacity of the proposed method of always providing a fast and feasible solution. With these results we pretend to remark the importance on the abstraction process that allowed us to reduce the classical MINLP problem into a set optimization allowing the integration of \textsf{FORWARD}.  Despite its superiority is not remarkable for smaller networks, the scalability properties of \textsf{FORWARD} method suppose significative advancement on time-consuming optimization processes. 

\subsection{Radial distribution with resource allocation selection}

As a natural step, we proceed by testing the proposed method over Problem $2$, which captures Problem $1$ as an specific instance. For this situation, where resource allocation corresponds to an optimization variable, Algorithm~\ref{alg:markov} has been compared against \textsf{KNITRO}, \textsf{JUNIPER} and a naive greedy search detailed previously.

In Table~\ref{tab:cost_b} and Table~\ref{tab:cpu_b} we can observe the load shed and the time required for finding a resource allocation and its radial distribution for each method. Similarly to previous section, it can be observed the improvement provided by Algorithm~\ref{alg:markov} among the other solvers with specific emphasis on the scalability properties. Besides, comparing it with the naive greedy search, it can be seen how the greedy approach is unable to find feasible solutions (NF) for some configurations as, conversely to Algorithm~\ref{alg:markov}, the greedy approach does not optimize considering feasibility which yields to inconsistent resource allocation. 

It is interesting to also highlight the memory allocation of the distinct methods, Table~\ref{tab:ram_b}. Branch-and-bound based methods tends to consider multiple combinations which exponentially grow if any feasible point is found, therefore, commercial solvers not only requires from powerful computers to solve the problems in an acceptable time, but also to be able to allocate all their internal variables. 

%%%%%%%%%%%%%%%%%% COST %%%%%%%%%%%%%%%%%%%
\begin{table}[t]
\centering
\caption{Distribution loss comparison (in kW) with constraint~\eqref{const::cardinality}. }
\label{tab:cost_b}
\begin{tabular}{|c|c|c|c|c|}
\hline
\textbf{Network} & \textbf{Permutation} & \textbf{Greedy} & \textbf{\textsf{KNITRO}} & \textbf{\textsf{JUNIPER}} \\ 
\hline
IEEE 11 & 284.257 & 284.257 & 284.257 & 284.257 \\
\hline
IEEE 18 & 175.821 & 175.821 & 204.328 & 69.197 \\
\hline
IEEE 33 & 325.402 & 285.821 & 285.821 & 285.821  \\
\hline
IEEE 37 & 472.485 & NF & 528.136 & TL  \\
\hline
IEEE 70 & 330.265 & 324.355 & TL & TL \\
\hline
\hline
IEEE 84 & 325.820 & 346.278 & TL & TL \\
\hline
IEEE 136 & 735.264 & 849.653 & TL & TL \\
\hline
IEEE 8500 & 1,121.65 & NF & TL &  TL \\
\hline
\hline
WS 120 & 1,428.72 & 1,568.25 & TL & TL \\
\hline
WS 240 & 3,425.82 & NF & TL & TL \\
\hline
WS 400 & 25,189.3 & 22,546.4 & TL & TL \\
\hline
\end{tabular}
\end{table}

%%%%%%%%%%%%%%%%%%%% TIME %%%%%%%%%%%%%%%%%%%
\begin{table}[t]
\centering
\caption{CPU time comparison when constraint~\eqref{const::cardinality}. }
\label{tab:cpu_b}
\begin{tabular}{|c|c|c|c|c|}
\hline
\textbf{Network} & \textbf{Permutation} & \textbf{Greedy} & \textbf{\textsf{KNITRO}} & \textbf{\textsf{JUNIPER}} \\ 
\hline
IEEE 11 & 1.40525 & 0.08178 & 14.8223 & 82.3094 \\
\hline
IEEE 18 & 1.12004 & 1.12004 & 194.467 & 423.782 \\
\hline
IEEE 33 & 0.22010 & 1.58800 & 5,267.92 & 26,892.31 \\
\hline
IEEE 37 & 2.43274 & NF & 6,648.27 & TL \\
\hline
IEEE 70 & 0.78306 & 3.82546 & TL & TL \\
\hline
\hline
IEEE 84 & 9.06977 & 19.8214 & TL & TL \\
\hline
IEEE 136 & 3.56218 & 14.1125 & TL & TL \\
\hline
IEEE 8500 & 15.8521 & NF & TL & TL \\
\hline
\hline
WS 120 & 3.98388 & 15.0016 & TL & TL \\
\hline
WS 240 & 4.26423 & NF & TL & TL \\
\hline
WS 400 & 8.78930 & 19.3415 & TL & TL \\
\hline
\end{tabular}
\end{table}

%%%%%%%%%%%%%%%%%%%%%%% RAM %%%%%%%%%%%%%%%%%%%%%
\begin{table}[t]
\centering
\caption{RAM allocation (in MiB) with constraint~\eqref{const::cardinality}.}
\label{tab:ram_b}
\begin{tabular}{|c|c|c|c|c|}
\hline
\textbf{Network} & \textbf{Permutation} & \textbf{Greedy} & \textbf{\textsf{KNITRO}} & \textbf{\textsf{JUNIPER}} \\ 
\hline
IEEE 11 & 180.691 & 2.620 & 1,851.46 & 3,226.10 \\
\hline
IEEE 18 & 160.193 & 2.432 & 865.931 & 917.624 \\
\hline
IEEE 33 & 5.234 & 53.301 & 4,487.49 & 5,832.35 \\
\hline
IEEE 37 & 68.029 & NF & 5,237.54 & TL \\
\hline
IEEE 70 & 17.356 & 83.809 & TL & TL \\
\hline
\hline
IEEE 84 & 195.898 & 402.456 & TL & TL \\
\hline
IEEE 136 & 236.784 & 473.901 & TL & TL \\
\hline
IEEE 8500 & 5,012.36 & NF & TL & TL \\
\hline
\hline
WS 120 & 232.700 & 375.981 & TL & TL \\
\hline
WS 240 & 658.431 & NF & TL & TL \\
\hline
WS 400 & 1,295.10 & 2,301.67 & TL & TL \\
\hline
\end{tabular}
\end{table}

\subsection{\textsf{FORWARD} as a warm start method}

Finally, as mentioned previously, the capacity of finding feasible solutions in short time of \textsf{FORWARD} method, and consequently of Algorithm~\ref{alg:markov}, yield us to analyze its possible usage as a warm-start strategy for the other solvers. Specifically, global solvers such as \textsf{KNITRO} relies part of its performance in finding a proper initial point. Actually, it has been demonstrated how these methods tends to fail, making impossible the obtainment of a solution. Therefore, in this section we integrated \textsf{FORWARD} method as a warm-start procedure for both \textsf{KNITRO} and \textsf{JUNIPER} solvers. 

\begin{figure}[t]
    \centering
    \includegraphics[width=0.45\textwidth]{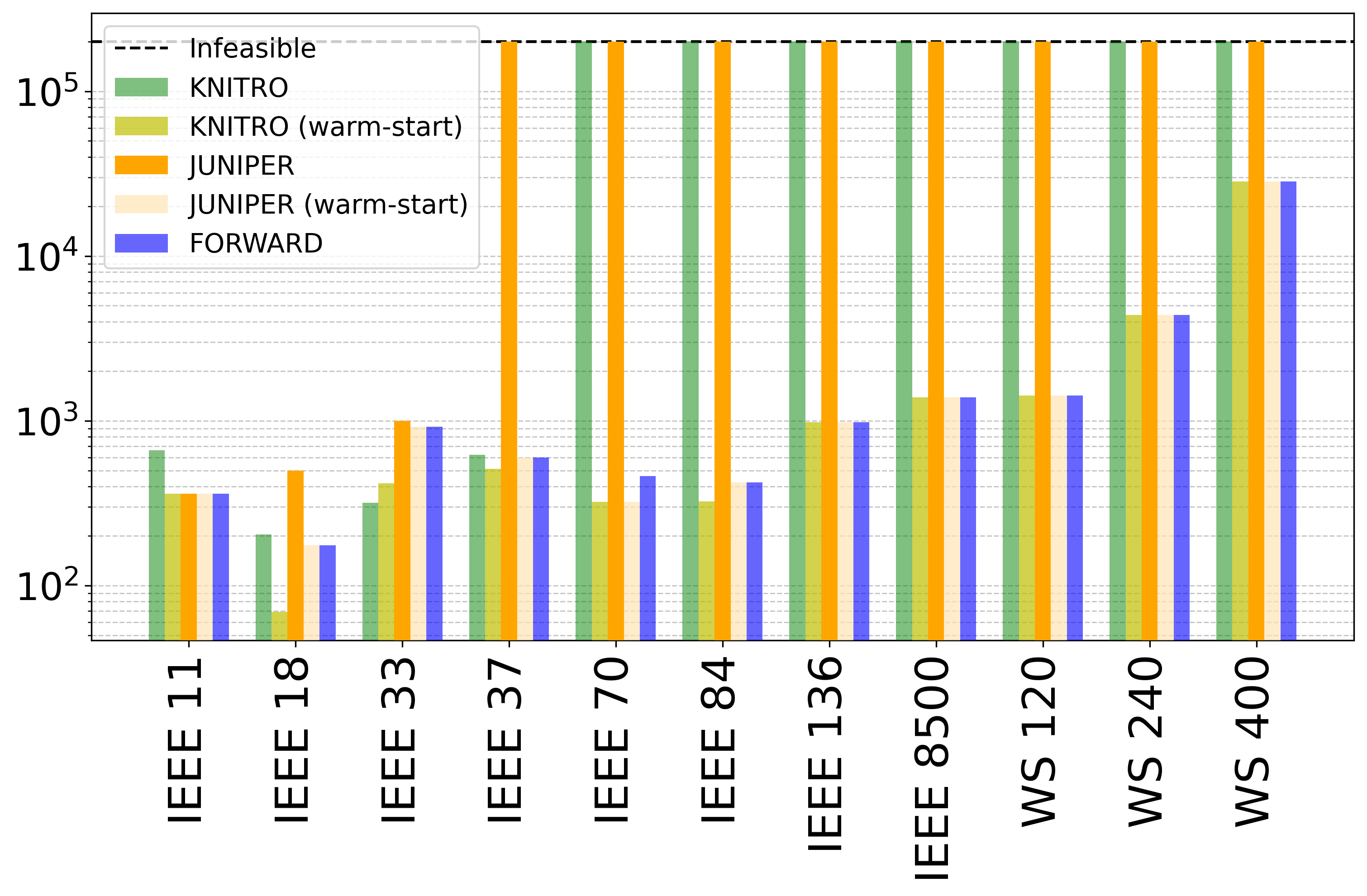} % Adjust the width as needed
    \caption{Loss of energy for each graph.}
    \label{fig:loss_plot}
\end{figure}

In Table~\ref{tab:cpu_c} it is clearly shown how \textsf{KNITRO} and \textsf{JUNIPER} solvers benefit from the \textsf{FORWARD} integration. Not only the computation time is substantially reduced with respect to their single implementation, Table~\ref{tab:cpu_a}, but also it increases its scalability and capacity to obtain solutions in acceptable time. Further comparison on the energy loss in each configuration can be observed in Fig.~\ref{fig:loss_plot}.
%and in Fig.~\ref{fig:percen_plot} where it is observed that the energy loss in the radial configurations never overpasses the $20\%$ as stated in Section~\ref{sec::intro}.

%%%%%%%%%%%%%%%%%%%% TIME %%%%%%%%%%%%%%%%%%%
\begin{table}[t]
\centering
\caption{CPU time comparison between Knitro, Juniper using \textsf{FORWARD} as a warm-start (in seconds). }
\label{tab:cpu_c}
\begin{tabular}{|c|c|c|}
\hline
\textbf{Network} & \textbf{\textsf{KNITRO}} & \textbf{\textsf{JUNIPER}}\\ 
\hline
IEEE 11 & 0.02362 & 71.3215\\
\hline
IEEE 18 & 1.72184 & 374.923\\
\hline
IEEE 33 & 1.40681 & 3,817.29 \\
\hline
IEEE 37 & 4.23834 & 57,918.7 \\
\hline
IEEE 70 & 8.41248 & 17,453.5 \\
\hline
\hline
IEEE 84 & 12.7581  & 25,129.4\\
\hline
IEEE 136 & 245.826 & 63,431.8\\
\hline
IEEE 8500 & 21,854.7 & 128,281 \\
\hline
\hline
WS 120 & 132.401 & 58,189.3 \\
\hline
WS 240 & 389.003 & 64,983.5 \\
\hline
WS 400 & 1,234.56 & 93,568.6 \\
\hline
\end{tabular}
\end{table}

%%%%%%%%%%%%%%%%%% PLOTS %%%%%%%%%%%%%%%%%%%%%%

\section{Conclusions}
\label{sec:conclusions}

We demonstrated \textsf{FORWARD} efficiency over real distribution problems by introducing the abstraction framework. It has been numerically proven the advantages of \textsf{FORWARD} over traditional MINLP solvers and shown that its integration as a warm-up strategy can be highly beneficial in terms of feasibility guarantees and computational speed. Finally, to solve the most general problem where active sources is also a decision variable, it has been also defined a Markov Chain search framework which offers probabilistic guarantees of optimality by keeping \textsf{FORWARD} as the core algorithm. Future work will focus on developing Markov Chain strategies for a tightest search with greater guarantees.

%%%%%%%%%%%% REFERENCES %%%%%%%%%%%%%%%%%
\bibliographystyle{ieeetr}
\bibliography{bib/alias,bib/reference,bib/joan}

\end{document}